\documentstyle[12pt]{article}
\begin{document}

\rightline{FTUV/93-28}

\begin{center}
{\LARGE{\bf Dynamical zeros in neutrino-electron elastic scattering at leading
 order}}
\end{center}

\vspace{2cm}

\begin{center}
{\large{\bf J. Segura, J. Bernab\'{e}u, F.J. Botella and J. Pe\~{n}arrocha }}

\vspace{1cm}

{\it Departament de F\'{\i}sica Te\'{o}rica \\ Universitat de Val\`{e}ncia \\
and \\ IFIC, Centre Mixte Univ. Valencia-CSIC \\ E-46100 Burjassot,
Spain}
\end{center}

\vspace{1.5cm}
\begin{center}
\end{center}
\vspace{2cm}

\begin{abstract}
\noindent
{\small{ We show the existence of dynamical zeros in the helicity amplitudes
for neutrino-electron elastic
scattering at lowest order in the standard theory.
In particular, the $\lambda=1/2$ non-flip electron helicity amplitude in the
electron antineutrino process vanishes for an incident neutrino energy
 $E_{\nu}=m_{e}/(4sin^{2}\theta_{W})$ and forward electrons (maximum
recoil energy). The rest of helicity amplitudes show kinematical zeros in this
configuration and therefore
the cross section vanishes. Prospects to search
for neutrino magnetic moment are discussed}}
\end{abstract}

\newpage

\baselineskip 0.7cm

The first $\nu_{i}(\bar{\nu}_{i}) e^{-}\rightarrow \nu_{i}(\bar{\nu}_{i})
e^{-}$ collision was observed in 1973 at Gargamelle
\cite{1}.In particular the observation of the process $\bar{\nu}_{\mu} e^{-}
\rightarrow \bar{\nu}_{\mu} e^{-}$ represented the discovery of neutral
currents, a milestone in the history of the standard model of electroweak
 interactions. In the year 1976 the
group of Reines \cite{2} got the first signal of the $\bar{\nu}_{e} e^{-}
\rightarrow \bar{\nu}_{e} e^{-}$ process, using neutrinos from a nuclear
 reactor.
Nowadays, there are also data from $\nu_{\mu} e^{-}
\rightarrow \nu_{\mu} e^{-}$ and $\nu_{e} e^{-}
\rightarrow \nu_{e} e^{-}$ reactions.

Needless to say, neutrino physics in general and these leptonic processes
in particular play a crucial role in the study of the standard model
of electroweak interactions, as well as in searching for effects beyond the
standard model:
 Charm
II collaboration has given values \cite{3}
 of the electroweak mixing
angle at a level of accuracy comparable with LEP data, the
( destructive ) interference among charged and neutral currents has been
measured \cite{4} in the reaction
$\nu_{e} e^{-}
\rightarrow \nu_{e} e^{-}$,
the laboratory bound  on the neutrino magnetic moment
($\mu_{\nu}<
2.4\small{x}10^{-10}$) has been set with $\bar{\nu}_{e} e^{-}
\rightarrow \bar{\nu}_{e} e^{-}$ \cite{2,5,6} in reactor experiments.
Besides, several new proposals  plan to reach
 a 1\% accuracy in the value of $sin^{2}\theta_{W}$ \cite{7},
 to study the $\bar{\nu}_{e}$ magnetic moment at the level of $2\small{x}
10^{-11}$
Bohr magnetons \cite{8} or even to search for flavour changing neutral
currents.

Let us concentrate in
the neutrino magnetic moment experiments.
The differential cross
section for $\bar{\nu}_{e} e^{-}
\rightarrow \bar{\nu}_{e} e^{-}$ including the neutrino magnetic moment
 contribution \cite{5} and neglecting
 neutrino mass
    is given by

\begin{equation}
\frac{\displaystyle{d\sigma_{\bar{\nu}_{e}}}}{\displaystyle{dT}}=
\frac{\displaystyle{2G^{2}m_{e}}}{\displaystyle{\pi}}\left[ g_{R}^{2}
+g_{L}^{2}\left(1-\frac{\displaystyle{T}}{\displaystyle{E_{\nu}}}\right)^{2}
-g_{L}g_{R}\frac{\displaystyle{m_{e}T}}{\displaystyle{E_{\nu}^{2}}}\right]
+\frac{\displaystyle{\pi \alpha^{2} \mu_{\nu}^{2}}}{\displaystyle{m_{e}^{2}}}
\frac{\displaystyle{(1-T/E_{\nu})}}{\displaystyle{T}}
\end{equation}
\

\noindent
where G is the Fermi coupling constant, $\alpha$ the fine structure
constant, $\mu_{\nu}$ the neutrino magnetic moment in Bohr magnetons,
$m_{e}$ the electron mass, T the   recoil kinetic
energy of the electron and
$E_{\nu}$ the antineutrino incident energy. In terms of
 $sin^{2}\theta_{W}$
the chiral couplings $g_{L}$ and $g_{R}$ can be written as

\begin{equation}
\begin{array}{cc}
g_{L}=\frac{1}{2}+sin^{2}\theta_{W}; & g_{R}=sin^{2}\theta_{W} \,\, .
\end{array}
\end{equation}
\

\noindent
 The first piece in the r.h.s of eq. (1) is the standard charged
and neutral current contribution ( for neutrinos one should exchange
$g_{L}$ by
 $g_{R}$ ), the second one is due to the neutrino magnetic moment, whose
 contribution adds incoherently.
  The magnetic moment contribution is enhanced with respect to the conventional
contribution for low values of T. Then, to extract information
about the neutrino magnetic moment term, it is extremely important
 to minimize the
experimental threshold on the electron recoil energy, which is a difficult
task. But from eq. (1), due to the presence of the
$g_{L}g_{R}$ term,  we observe that a different strategy can be used
, namely, to look
for kinematical configurations, if any,  where the standard model
 contribution vanishes. Instead of looking for regions where the new physics
becomes large enough to be comparable to the standard contribution we will
 look for configurations were the standard contribution becomes small
enough.

With this motivation , we study in this paper the dynamical zeros
of the helicity
 amplitudes for neutrino and antineutrino scattering with electrons at lowest
order in the standard model. By dynamical zeros we mean the ones which
appear inside the physical region of the kinematical variables describing
the scattering \cite{9}. The location of these dynamical zeros
 depends on the values of
the fundamental parameters of the electroweak theory, namely, $g_{L}$ and
$g_{R}$ in our case. Besides dynamical zeros, the helicity amplitudes are
constrained by the kinematical zeros \cite{10} which appear at the boundary
of the physical region and do not depend on dynamical parameters.

On searching for dynamical zeros in neutrino and antineutrino electron
 scattering
one can make a first attempt by selecting a kinematical configuration where
only one spin amplitude contributes to the standard model cross section
in eq. (1). In particular, in a general collinear frame the helicity
amplitude $M_{\lambda ' \lambda }^{\nu_{i}}$ for \mbox{$\nu_{i} e^{-}
\rightarrow \nu_{i} e^{-}$}, $i=e,\mu,\tau$,
 with $\lambda =\lambda '=-1/2 \,$ is the only one
 that
contributes for backward outgoing neutrino  ( forward electron ), being
$\lambda$ and $\lambda '$ the initial and final electron
 helicities respectively.
This follows from angular momentum conservation arguments.
 For  $\bar{\nu}_{i} e^{-}
\rightarrow \bar{\nu}_{i} e^{-}$ the helicity amplitude $M_{1/2,1/2}^
{\bar{\nu}_{i}} \, \,$ is the only one that contributes for backward
 neutrino.

  In the LAB frame, the backward cross section for $\nu_{i} e^{-}
\rightarrow \nu_{i} e^{-}$
 can be written as follows

\begin{equation}
\left(\frac{\displaystyle{d\sigma_{\nu_{i}}}}{\displaystyle{dT}}\right)_{back}
=\frac{\displaystyle{2G^{2}m_{e}}}{\displaystyle{\pi}}
\left[ g^{i}_{L}-g^{i}_{R}\frac{\displaystyle{m_{e}}}{\displaystyle{
2E_{\nu}+m_{e}}}\right]^{2}
\end{equation}
\

\noindent
which is proportional to $\mid M_{-1/2,-1/2}^{\nu_{i}}
(back)\mid^{2}$.
Eq. (3) is easily obtained from the standard piece of eq. (1) considering
that the value of T in the backward configuration is
\begin{equation}
T_{max}=\frac{\displaystyle{2E_{\nu}^{2}}}{\displaystyle{2E_{\nu}+m_{e}}}.
\end{equation}
\

 In eq. (3) there are no dynamical zeros for this backward cross
section with $g_{L}^{e}$ and $g_{R}^{e}$ satisfying $g_{L}^{e}>g_{R}^{e}>0$,
which is the
case for $\nu_{e}$ as seen from equation (2). On the other hand, for
$\bar{\nu}_{i} e^{-}$ backward elastic scattering we have

\begin{equation}
\left(\frac{\displaystyle{d\sigma_{\bar{\nu}_{i}}}}{\displaystyle{dT}}
\right)_{back}
=\frac{\displaystyle{2G^{2}m_{e}}}{\displaystyle{\pi}}
\left[ g^{i}_{R}-g^{i}_{L}\frac{\displaystyle{m_{e}}}{\displaystyle{
2E_{\nu}+m_{e}}}\right]^{2}
\end{equation}
\

\noindent
which is proportional to $\mid M_{1/2,1/2}^{\bar{\nu}_{i}}
(back)
\mid^{2}$ and vanishes for $\bar{\nu}_{e}$ at

\begin{equation}
E_{\nu}=
\frac{\displaystyle{m_{e}}}{\displaystyle{4sin^{2}\theta_{W}}}\, \, .
\end{equation}
\

\noindent
Therefore we have found that for the  antineutrino energy $E_{\nu}$ given
by equation (6) and the corresponding maximum electron recoil energy given
by equation (4) the
differential cross section for $\bar{\nu}_{e}e^{-}\rightarrow \bar{\nu}_{e}
e^{-}$
vanishes exactly at leading order.
It is worthwhile to emphasize that $E_{\nu}\simeq m_{e}$ lies inside
the range of reactor antineutrino spectra and
 $T=T_{max}\simeq \frac{2}{3}m_{e}$ is in the range of the proposed
 experiments to
detect recoil electrons \cite{8}.

For $\nu_{\mu}$ and $\bar{\nu}_{\mu}$ elastic scattering ( or
$\nu_{\tau}$ and $\bar{\nu}_{\tau}$ ) the corresponding $g_{L}^{\mu}$,
 $g_{R}^{\mu}$ parameters are

\begin{equation}
\begin{array}{cc}
g_{L}^{\mu}=-\frac{1}{2} +sin^{2}\theta_{W}; &
g_{R}^{\mu}=sin^{2}\theta_{W}
\end{array}
\end{equation}
\

These values prevent the corresponding cross sections from dynamical zeros
for backward neutrinos. It is therefore evident that the contribution
of charged currents to the values of $g_{L}^{i}$ and
$g_{R}^{i}$ is essential in the existence of the dynamical zeros given
 by eqs. (4) and (6) in $\bar{\nu}_{e} e^{-}$ scattering.

In what follows we will present a systematic analysis of the dynamical zeros
 of the helicity amplitudes for neutrino and antineutrino - electron scattering
at lowest order in electroweak interactions. As a consequence of this analysis
one obtains all the information about dynamical zeros for polarized and
unpolarized differential cross sections.

For $\nu_{i} e^{-}\rightarrow \nu_{i} e^{-}$ the helicity amplitudes are

\begin{equation}
M_{\lambda ',\lambda}^{\nu_{i}}=\frac{\displaystyle{G}}{\displaystyle{
\sqrt{2}}}(g_{L}^{i} M_{L}^{\nu_{i}}+g_{R}^{i}M_{R}^{\nu_{i}})
\end{equation}
\

\noindent
where

\begin{equation}
\begin{array}{c}
M_{L}^{\nu_{i}}=[\bar{u}_{e}'(\lambda ')\gamma_{\alpha}(1-\gamma_{5})
u_{e}(\lambda )][\bar{u}'_{\nu_{i}}\gamma^{\alpha}(1-\gamma_{5})
u_{\nu_{i}}]\\
M_{R}^{\nu_{i}}
=[\bar{u}_{e}'(\lambda ')\gamma_{\alpha}(1+\gamma_{5})
u_{e}(\lambda )][\bar{u}'_{\nu_{i}}\gamma^{\alpha}(1-\gamma_{5})
u_{\nu_{i}}]
\end{array}
\end{equation}
\

\noindent
We carry out the explicit calculation in the LAB frame, by using a standard
representation of the Dirac matrices and spinors \cite{11}.
The final result for the helicity amplitudes reads as follows

\begin{equation}
\begin{array}{l}
M^{\nu_{i}}_{+,+}=Ng_{R}^{i}\left( 1+\frac{\displaystyle{T}}
{\displaystyle{\mid\vec{p'}\mid }}\right)
sin\frac{\displaystyle{\theta}}{\displaystyle{2}}
cos\frac{\displaystyle{\beta}}{\displaystyle{2}}\\
\\
M^{\nu_{i}}_{-,-}=N\left[
g_{L}^{i}\left( 1+\frac{\displaystyle{T}}
{\displaystyle{\mid\vec{p'}\mid }}\right)
sin\frac{\displaystyle{\theta - \beta}}{\displaystyle{2}}
+
g_{R}^{i}\left( 1-\frac{\displaystyle{T}}
{\displaystyle{\mid\vec{p'}\mid }}\right)
cos\frac{\displaystyle{\theta}}{\displaystyle{2}}
sin\frac{\displaystyle{\beta}}{\displaystyle{2}}\right]\\
\\
M^{\nu_{i}}_{+,-}=N\left[
-g_{L}^{i}\left( 1-\frac{\displaystyle{T}}
{\displaystyle{\mid\vec{p'}\mid }}\right)
cos\frac{\displaystyle{\theta - \beta}}{\displaystyle{2}}
+
g_{R}^{i}\left( 1+\frac{\displaystyle{T}}
{\displaystyle{\mid\vec{p'}\mid }}\right)
sin\frac{\displaystyle{\theta}}{\displaystyle{2}}
sin\frac{\displaystyle{\beta}}{\displaystyle{2}}\right]\\
\\
M^{\nu_{i}}_{-,+}=Ng_{R}^{i}\left( 1-\frac{\displaystyle{T}}
{\displaystyle{\mid\vec{p'}\mid }}\right)
cos\frac{\displaystyle{\theta}}{\displaystyle{2}}
cos\frac{\displaystyle{\beta}}{\displaystyle{2}}
\end{array}
\end{equation}
\

\noindent
where $N=8G\sqrt{E_{\nu} E_{\nu} ' m_{e} (T+2m_{e})}$, $\mid\vec{p'}\mid$
 is the outgoing electron momentum and $\theta$ and $\beta$ are the
counterclockwise angles in the scattering plane of the final electron and
 neutrino with respect
to the incoming neutrino direction.
We note
that the helicity  of the target electron ( at rest ) is referred to the
backward direction. In eq. (10) we have used the helicity signs instead of
its values.

We have the following relation between neutrino and antineutrino helicity
amplitudes

\begin{equation}
M_{\lambda ',\lambda}^{\bar{\nu}_{i}}(g_{L}^{i},g_{R}^{i})=
(-)^{\lambda '-\lambda}M_{-\lambda ',-\lambda}^{\nu_{i}}
(g_{R}^{i},g_{L}^{i})
\end{equation}
\

\noindent
i.e., apart from the phase factor $(-)^{\lambda '- \lambda}$, one has
to replace $g_{R}^{i}\leftrightarrow g_{L}^{i}$ and change the sign of
helicities going from neutrino to antineutrino amplitudes.

{}From eqs. (10) and (11) it is clear that the amplitudes $M_{\pm ,-}^
{\nu_{i}}$ and $M_{\pm ,+}^
{\bar{\nu}_{i}}$ are the only ones that get contribution from both
 $M_{L}$ and $M_{R}$. Therefore, these
are the only amplitudes which may have dynamical zeros, while the others will
not exhibit any unless $g_{R}^{i}$ or $g_{L}^{i}$ vanish. The
 conditions that define the dynamical zeros for the
helicity amplitudes are the following:

\begin{equation}
\begin{array}{ccc}
M_{\pm ,-}^{\nu_{i}}=0 & \leftrightarrow &
cos\theta =\mp \frac{\displaystyle{m_{e}+E_{\nu}}}{\displaystyle{m_{e}
\frac{\displaystyle{g_{R}^{i}}}{\displaystyle{g_{L}^{i}}}
-E_{\nu}}}
\end{array}
\end{equation}
\

\noindent
and

\begin{equation}
\begin{array}{ccc}
M_{\pm ,+}^{\bar{\nu}_{i}}=0 & \leftrightarrow &
cos\theta =\pm \frac{\displaystyle{m_{e}+E_{\nu}}}{\displaystyle{m_{e}
\frac{\displaystyle{g_{L}^{i}}}{\displaystyle{g_{R}^{i}}}
-E_{\nu}}}
\end{array}
\end{equation}
\

Taking into account that the physical region is restricted by $0\leq cos\theta
\leq 1$, and the $g_{L}^{i}$ and $g_{R}^{i}$ values it is straightforward
 to arrive
to the following conclusions:

{\bf i)} $M_{++}^{\bar{\nu}_{e}}$ shows dynamical zeros given by eq. (13) in
the
energy range

\noindent
 \mbox{$0\leq E_{\nu}\leq m_{e}/4sin^{2}\theta_{W}$}.
 The upper value
 corresponds to the phase space point \mbox{$cos\theta =1$}.
At this end point the
 other
three helicity amplitudes have  kinematical zeros as can be
explicitly seen from eq. (10). This is the reason why
this dynamical zero shows up in the unpolarized cross section in the
backward configuration as we already pointed out in eqs. (5) and (6).

{\bf ii)} $M_{-+}^{\bar{\nu}_{\mu},\bar{\nu}_{\tau}}$ show dynamical zeros
given by eq. (13) in the whole range of energies $0\leq E_{\nu}<\infty$ .
 In this case the
helicity amplitudes never vanish simultaneously. Then, the dynamical
 zeros will only show up in polarized cross sections.

{\bf iii)} There are no more solutions of eqs. (12) and (13) in the
physical region.\\

These results are summarized in figure 1, where the dynamical zeros are plotted
in the
 plane $(E_{\nu},cos\theta)$ , together with the kinematical zeros.

It seems difficult to design a $\bar{\nu_{e}} e^{-}$ experiment where electron
polarizations are involved. So we shall concentrate in the dynamical zero
defined by eqs. (4) and (6); the only one we consider relevant for
realistic experimental proposals.

The fact that the weak backward cross section for
$\bar{\nu_{e}} e^{-}\rightarrow
\bar{\nu_{e}} e^{-}$ vanishes at leading order for $E_{\nu}=
m_{e}/(4sin^{2}\theta_{W})$ clearly points out that this kinematical
 configuration must be a good
place to study new physics or even higher order electroweak
radiative corrections \cite{12}. It is worthwhile
to emphasize once again that backward neutrinos  correspond to forward
electrons
 with maximum recoil kinetic  energy which is a very interesting situation
from the experimental point of view. This configuration, showing the
dynamical zero, is of particular interest to look for those contributions
which add incoherently to the standard amplitude.

To illustrate the interest of this dynamical zero we shall
concentrate in the possibility of searching for neutrino magnetic moment
. In Figure 2 we denote by
$(d\sigma_{W}/dT)_
{back}$ the standard contribution in the r.h.s. of eq. (1)
and by $(d\sigma_{M}/dT)_{back}$
the magnetic moment contribution
 , both for $T=T_{max}$.
The solid line represents the boundary where
$(d\sigma_{W}/dT)_
{back}=(d\sigma_{M}/dT)_
{back}$ for $\bar{\nu_{e}} e^{-}\rightarrow
\bar{\nu_{e}} e^{-}$. The regions below the other lines are those for which
$(d\sigma_{M}/dT)_
{back}>(d\sigma_{W}/dT)_
{back}$
 for the rest of neutrino species. It is quite apparent from this
figure that electron antineutrinos with energies around 0.5 MeV give the
possibility of studying low values for neutrino magnetic moment. With other
kind of neutrinos this is only possible by going to much lower values of
neutrino energy.

In conclusion we have discussed all the dynamical zeros in the
helicity amplitudes for neutrino (antineutrino)-electron scattering.
Particularly interesting is the electron antineutrino backward configuration,
where the only allowed helicity amplitude has a dynamical zero for
$E_{\nu}=m_e/(4sin^{2}\theta_{W})$, so the backward unpolarized cross section
 vanishes
 at lowest order in the standard theory.
This result clearly points out a favourable kinematical configuration to
look for new physics in $\bar{\nu_{e}} e^{-}\rightarrow \bar{\nu_{e}}
e^{-}$. Results have been presented for the expectations
to search for neutrino magnetic moment.
\vspace{2cm}

{\bf ACKNOWLEDGEMENTS}

\hspace{0.5cm}This paper has been supported by CICYT under Grant AEN 90-0040
. J.S. thanks the Spanish Ministry of Education for his fellowship. We are
 indebted to Prof. S. Bilenky and Prof. F. Halzen
for discussions about the topic of this paper.

\newpage
\begin{center}
{\large {\bf Figure Captions.}}
\end{center}

\begin{itemize}
     \item{{\bf Fig. 1)} Kinematical and dynamical zeros for the helicity
amplitudes in the plane ($E_{\nu},cos\theta$).}
     \item{{\bf Fig. 2)} Regions of dominance of weak or magnetic backward
 differential cross
 sections in the
 plane ($\mu_{\nu},E_{\nu}$)
for $\bar{\nu}_{e}$; there are three different zones divided by the
solid line. For the rest of (anti-)neutrinos there are only two regions, being
the magnetic backward cross section dominant above the corresponding line
( long-dashed for $\nu_{e}$, dashed-dotted for $\bar{\nu}_{\mu}$ and
short-dashed for $\nu_{\mu}$ ) and the opposite below the line.}
\end{itemize}

\end{document}